\documentclass[aip,amsmath,amssymb,reprint,onecolumn, nofootinbib]{revtex4-1}

\usepackage{graphicx}
\usepackage{dcolumn}
\usepackage{bm}
\usepackage{xcolor}
\usepackage[utf8]{inputenc}
\usepackage[T1]{fontenc}
\usepackage{mathptmx}
\usepackage{etoolbox}

\makeatletter
\def\@email#1#2{
 \endgroup
 \patchcmd{\titleblock@produce}
  {\frontmatter@RRAPformat}
  {\frontmatter@RRAPformat{\produce@RRAP{*#1\href{mailto:#2}{#2}}}\frontmatter@RRAPformat}
  {}{}
}

\definecolor{MyB}{rgb}{0.1,0.1,1.0}

\usepackage{hyperref}
\usepackage{changes}
\usepackage[section]{placeins}
\usepackage[titletoc]{appendix}
\usepackage{xcolor}
\usepackage{dsfont}
\usepackage{bm}
\usepackage{mathtools}
\usepackage{enumerate}

\definecolor{NewOrange}{rgb}{0.800, 0.250, 0.100}

\def\tC{\tau}
\def\tQN{t}

\DeclareMathAlphabet{\mathpzc}{OT1}{pzc}{m}{it}

\newcommand{\sayy}[1]{`#1'}
\DeclarePairedDelimiter\abs{\lvert}{\rvert}
\usepackage{calc}
\usepackage{accents}

\providecommand{\href}[2]{#2}

\usepackage{amsthm}

\def\be{\begin{equation}}
\def\ee{\end{equation}}
\def\bea{\begin{eqnarray}}
\def\eea{\end{eqnarray}}

\def\sig{\sigma}
\def\hsig{\hat{\sigma}}

\def\la{\langle}
\def\ra{\rangle}

\def\obs{\mathcal{O}}

\makeatother
\begin{document}
\title{Observational Quantities in Quasi-Newtonian Descriptions of Cosmological Space-Times \phantom{aaaaaaaaaaaaaaaaaaaaaaaaaaaaaaaaaaaaaaaaaaaaaaaaaaaaaaaaaaaaaaaaaaa}} 

\author{Asta~Heinesen$^*$}
\email{asta.heinesen@nbi.ku.dk}
 \affiliation{Niels Bohr Institute, Blegdamsvej 17, DK-2100 
Copenhagen, Denmark} 
 \affiliation{Department of Physics and Astronomy, Queen Mary University of London, UK}

\author{Davide~Fontana}
 \affiliation{Department of Physics and Astronomy, Queen Mary University of London, UK}

\author{Timothy~Clifton}
 \affiliation{Department of Physics and Astronomy, Queen Mary University of London, UK}

\begin{abstract} 
We investigate measures of distance and redshift in cosmological space-times that admit a shear-free foliation, which we henceforth refer to as `quasi-Newtonian'. Space expands isotropically in this description, and small-scale gravitational physics has a natural Newtonian limit, which makes it ideal for considering the physics of wide classes of cosmological models. By assuming that the energy-momentum tensor is dominated by rest-mass density, and that the 3-velocity of matter is small in the quasi-Newtonian frame, we derive fundamental results for kinematics and light propagation. Our results provide a new way of formulating general-relativistic cosmologies with non-perturbative structures in terms of quantities that can be understood from cosmological perturbation theory and post-Newtonian expansions, and allow us to quantify departures of observables from the predictions of Friedmann cosmology. It thereby provides a route to understanding inherently relativistic space-time structures, such as those that occur in Lema\^{i}tre-Tolman-Bondi,  Szekeres solutions, and Bianchi cosmologies in terms of Newtonian degrees of freedom. We illustrate our results using the degenerate Kasner solution as an example, and explain how our approach can be used to provide new insights into the current cosmological tensions.

\end{abstract}

\maketitle

\section{Introduction}
\label{sec:intro} 

Statistically homogeneous relativistic cosmological models can be constructed using a range of methods; from arranging black holes in a lattice \cite{lattice}, to constructing Swiss cheese space-times containing regions of Lema\^{i}tre-Tolman-Bondi (LTB) \cite{clifton2009hubble, fleury2014swiss} and/or Szekeres solutions \cite{Lavinto:2013exa, Koksbang:2017arw}, and by using post-Newtonian methods \cite{sanghai2017ray} and numerical simulations \cite{Adamek:2018rru, Macpherson:2022eve}. Rather surprisingly, these models often have optical properties that conform reasonably well to the distance-redshift relations derived in the perfectly homogeneous and isotropic Friedmann-Lema\^{i}tre-Robertson-Walker (FLRW) models. This could be due to the adoption of large-scale symmetries that usually occur within them, either through the imposition of a background or through initial conditions or an assumption of spatial periodicity. Even so, it is still not \emph{a priori} clear why the FLRW observational relations perform quite so well, and why there is no accumulation of non-linear effects along beams of light in these less symmetric approaches. After all, statistically homogeneous cosmologies often have kinematics that deviate substantially from those of a perfect Friedmann universe, with non-constant and anisotropic rates of expansion of matter world-lines, and which superficially obey equations that can look quite different to those of the FLRW class of cosmologies.

We address this problem by following the approach of van~Elst and Ellis, in which they re-foliate space-time in order to find a description in which links to Friedmannian cosmology and Newtonian descriptions of gravity can be made \cite{vanElst:1998kb}. This results in a {\it quasi-Newtonian} description of cosmology, which shares some of the properties of the Poisson gauge of the scalar sector of cosmological perturbation theory  around an FLRW background; it has time-like normals that are both shear and vorticity-free, and it naturally leads to the Newton-Poisson equation describing the leading-order part of the motion of pressureless matter fields. 
The great benefit of the quasi-Newtonian approach, however, is that it does {\it not} require one to make any assumptions about global symmetries of the Universe, beyond the fact that the space-time should admit the required foliation. It therefore allows a natural generalization of the Poisson gauge beyond the regime of FLRW cosmology, while simultaneously allowing us to identify the aspects of non-FLRW cosmologies that could produce new and interesting observations.

We will start our analysis by recapping some of the essential physics from the quasi-Newtonian approach to cosmology, before specializing our considerations to cosmologies in which the energy-momentum tensor is dominated by the rest mass of the matter fields within it, and in which the 3-velocity of matter is small in the preferred foliation. We will then calculate observables in these space-times in terms of the variables used in the quasi-Newtonian description, and comment on the degree to which they resemble those of perturbed FLRW models. This will include measures of distance and redshift. We will then consider the degenerate Kasner solution as a simple example space-time in which we can explicitly demonstrate the validity of our results for observational quantities, as well as consider the ways in which this description indicates that we could generalize existing approaches to modeling and simulating the Universe to include more general cosmological behaviors. We use units where $c=1$, and use Greek letters $\mu, \nu, \ldots$ for space-time indices of tensors in a general coordinate basis, while lowercase Latin letters $i,j,\ldots$ denote spatial indices only.

\section{Quasi-Newtonian space-times}
\label{sec:QN} 
Following van Elst and Ellis \cite{vanElst:1998kb}, we define a \emph{quasi-Newtonian} space-time as being one that admits a shear-free foliation. That is, we assume that there exists a set of connected 3-spaces with time-like normal, $n_{\mu}$, that can be kinematically decomposed as
\bea
\label{def:nablan}
\nabla_{\nu}n_\mu  = \frac{1}{3}\theta h_{\mu \nu } +\sig_{\mu \nu} + \omega_{\mu \nu}  - n_\nu a_\mu \, ,
\eea 
where $a_{\mu} \coloneqq n^\nu \nabla_\nu n_\mu$ is the 4-acceleration of an observer following $n^{\mu}$, $\theta \coloneqq \nabla_{\mu}n^{\mu}$ is the isotropic rate expansion of the space orthogonal to $n^{\mu}$, and where the shear and vorticity are given by
\bea
\label{QNdef}
\sig_{\mu \nu} \coloneqq h_{  \la \nu  }^{\, \beta}  h_{ \mu \ra }^{\, \alpha }\nabla_{  \beta}n_{\alpha  } =0 \quad {\rm and} \quad  \omega_{\mu \nu} \coloneqq h_{  [\nu  }^{\, \beta}  h_{  \mu] }^{\, \alpha }\nabla_{  \beta}n_{\alpha  } =0   \,  ,
\eea
where $h_{\mu \nu} \coloneqq g_{\mu \nu} + n_{\mu} n_{\nu}$ is the induced metric on the 3-space orthogonal to $n^{\mu}$. The angular brackets denote the projected-symmetric and trace-free part of a tensor round brackets denote symmetrization and square brackets denote anti-symmetrization, such that e.g. $S_{\la \mu\nu \ra} \coloneqq (h_{ \mu  }^{\;\;\; \alpha}h_{ \nu  }^{\;\;\; \beta}  - \frac{1}{3}h_{ \mu  \nu}h^{\alpha\beta})S_{(\alpha\beta)}$ and $S_{(\mu\nu)}\coloneqq \frac{1}{2}(S_{\mu\nu}+S_{\nu\mu})$ and $S_{[\mu\nu]}\coloneqq \frac{1}{2}(S_{\mu\nu} - S_{\nu\mu})$. The vanishing of $\sig_{\mu \nu}$ is the shear-free condition required for the space to be expanding isotropically, while the vanishing of $\omega_{\mu \nu}$ is necessary and sufficient for the spaces orthogonal to $n^{\mu}$ to be hypersurface-forming, as it implies that $n_\mu$ can be written as a gradient:
\bea
\label{gradient} 
\hspace*{-0.6cm}  n_\mu  = - N \nabla_\mu \tQN \, ,
\eea 
where $\tQN$ is a global time parameter that specifies the leaf of the foliation, and the \emph{lapse} function $N=(- g^{\mu \nu} \nabla_\mu \tQN \nabla_\nu \tQN )^{-1/2}$ ensures the normalization $n^\mu n_\mu = -1$. We can then identify the 4-acceleration of an observer following an integral curve of $n^{\mu}$ as
\bea
\label{vorticitydot}
\hspace*{-0.8cm}  
a_\mu = D_\mu \ln(N) \, , 
\eea 
where {$D_{\mu} S^{\alpha \ldots}_{\;\;\;\beta \ldots} \coloneqq  h^{\alpha}_{\;
\;\;\gamma}\, \ldots h_{\beta}^{\;\;\;\delta} \ldots h_{\mu}^{\;\;\;\nu}\nabla_{\nu}S^{\gamma\ldots}_{\;\;\;\delta\ldots}$} is a spatially-projected derivative for a tensor $S^{\alpha \ldots}_{\;\;\;\beta \ldots}$ of arbitrary rank.  From this, the electric and magnetic parts of the Weyl tensor $C_{\alpha\beta\mu\nu}$ are {{defined as}
\bea
\label{sheardot1}
\hspace*{-0.8cm}  
&E_{\mu \nu }& \coloneqq n^\rho n^\sigma  C_{\rho \mu \sigma \nu}  = D_{\la \mu} a_{\nu \ra}  + a_{\la \mu} a_{\nu \ra}    \\ \label{magnetic}
&H_{\alpha\beta}& \coloneqq
-\frac{1}{2}\epsilon_{\rho\sigma\gamma\delta}
C_{\mu\nu}{}^{\gamma\delta}n^{\rho}h_{\alpha}{}^{\sigma}
n^{\mu}h_{\beta}{}^{\nu}  =  0 \, ,
\eea 
with $\epsilon_{\alpha\beta\mu\nu}$ being the totally antisymmetric {Levi-Cevita} tensor. Invoking the Einstein Equations and using the embedding equations for the 3-dimensional spatial Ricci tensor of the quasi-Newtonian surfaces, we have that the vanishing shear condition implies  
\bea
\label{Rtrace2} 
\hspace*{-0.3cm}  {}^{(3)}\! R  &=&  16\pi G T_{\mu \nu} n^\mu n^\nu - \frac{2}{3} \theta^2 + 2\Lambda \, ,
\eea 
and 
\bea
\label{spatialRicci1}
\hspace*{-0.8cm}  
{}^{(3)}\! R_{\la \mu \nu \ra} =   E_{\mu \nu }     \, , 
\eea 
where $T_{\mu \nu}$ is the energy-momentum tensor and $\Lambda$ is the cosmological constant. The vanishing shear condition further implies 
\bea
\label{def:lie}
&&  \pounds_{\bm n} h_{\mu \nu } \coloneqq  2 h^{\alpha}_{\nu} h^{\beta}_{\mu} \nabla_{(\alpha} n_{\beta)}  =  \frac{2}{3}\theta h_{\mu \nu }  \, .
\eea 
 This means that the lie transport of the spatial metric tensor $h_{ \mu \nu }$ along the flow lines of $n^{\mu}$ is given by a conformal mapping, such that
\bea
\label{def:hconformal}
&&   h_{\mu \nu }  = \left( \frac{\mu }{ \mu_{\text{i}}} \right)^{\frac{2}{3}}  \hat{h}_{\mu \nu}  \,   \qquad {\rm and} \qquad  \pounds_{\bm n} \hat{h}_{\mu \nu } = 0 \, ,
\eea 
where $\mu \coloneqq \sqrt{\det (h_{ij})}(\tQN,x^i)$ is the volume density in adapted coordinate defined by $n^\mu \nabla_\mu x^i = 0$, and $\mu_{\text{i}}$ is the value of $\mu$ at some initial time $\tQN=\tQN_{\rm i}$. The re-scaled metric $\hat{h}_{\mu \nu }$ is then equal to $h_{\mu \nu}$ as time $\tQN=\tQN_{\text{i}}$, such that 
\bea
\label{def:muint}
\hspace{-0.5cm}  \mathfrak{a} \coloneqq \frac{\mu^\frac{1}{3} }{ \mu_{\text{i}}^\frac{1}{3} }  &=&   \exp{\left(  \frac{1}{3} \int_{ \tQN_{\text{i}} }^{\tQN} \!\!\!\! {\rm d} \tQN'    N \theta \right)} \,  , 
\eea 
where the integral is performed along the flow lines of $n^{\mu}$. We can now write the trace and trace-free parts of three-dimensional Ricci tensor ${}^{(3)}\! R_{\mu \nu}$ in terms of the ${}^{(3)}\! \hat{R}_{\mu \nu}$ associated with the rescaled metric $\hat{h}_{\mu \nu }$ and the scaling factor $\mu$: 
\bea
\label{Rtrace} 
\hspace*{-0.3cm}  {}^{(3)}\! R  =&&  \frac{1}{\mathfrak{a}^2} {}^{(3)}\! \hat{R}  - \frac{4}{\mathfrak{a}}   h^{\mu \nu} D_{\mu} D_{\nu} \mathfrak{a}  +    \frac{6}{\mathfrak{a}^2}  h^{\mu \nu} D_{\mu} \mathfrak{a} \, D_{\nu} \mathfrak{a}   \, , 
\eea 
and 
\bea
\label{Rtracefree} 
\hspace*{-0.5cm}  {}^{(3)}\! R_{\la \mu \nu \ra} &=&   {}^{(3)}\! \hat{R}_{\la \mu \nu \ra}  -    \frac{1}{\mathfrak{a}} D_{\la \mu} D_{\nu \ra} \mathfrak{a} \, , 
\eea 
respectively. In the special case where $\mu$ is homogeneous over the hypersurfaces defined by $n^\mu$, only the first term of Equation \eqref{Rtrace} remains. This gives ${}^{(3)}\! R \propto  \mathfrak{a}^{-2}$, which resembles the Friedmannian scaling law for curvature if $\mathfrak{a}$ were replaced by the FLRW scale factor. 

In what follows, we will be primarily interested in modeling the largest scales in the Universe, where matter is expected to be well-described as dust. We thus require that there exists a set of flow lines generated by a time-like 4-velocity field $u^\mu$, such that the energy-momentum tensor takes the form
\bea
\label{T}
T_{ \mu  \nu } = u_{ \mu } u_{\nu } \rho \, , 
\eea  
where $\rho \coloneqq T_{ \mu  \nu } u^{\mu} u^{\nu}$ is the rest-mass density of the dust. The general kinematical decomposition of $u^\mu$ is then given by
\bea
\label{def:nablau}
&& \nabla_{\nu}u_\mu  = \frac{1}{3} \tilde{\theta} \tilde{h}_{\mu \nu }   +\tilde{\sig}_{\mu \nu} + \tilde{\omega}_{\mu \nu}  
\eea 
where the expansion, shear and vorticity of the dust in its own rest-frame are given by $\tilde{\theta} \coloneqq \nabla_{\mu}u^{\mu}$, $\tilde{\sig}_{\mu \nu} \coloneqq \tilde{h}_{  \la \nu  }^{\, \beta}  \tilde{h}_{ \mu \ra }^{\, \alpha }\nabla_{  \beta}u_{\alpha  }$ and $\tilde{\omega}_{\mu \nu} \coloneqq \tilde{h}_{  [\nu  }^{\, \beta}  \tilde{h}_{  \mu] }^{\, \alpha }\nabla_{  \beta}u_{\alpha  }$, respectively. Here, we have used $u^\nu \nabla_\nu u^\mu = 0$, which follows directly from the form of the energy-momentum tensor \eqref{T} and its conservation.

The 4-velocity of the dust, $u^{\mu}$, can now be written in terms of the time-like normals to the quasi-Newtonian foliation, $n^{\mu}$, as
\bea
\label{udecomp}
 u^\mu = \gamma (n^\mu + v^\mu) \, , 
 \eea
where $\gamma \coloneqq - n^\mu u_\mu =(1 - v^\mu v_\mu )^{- \frac{1}{2}}$ is the Lorentz factor between the two frames defined by $u^{\mu}$ and $n^{\mu}$, and $v^\mu$ the \emph{peculiar velocity field} of the matter in the quasi-Newtonian rest frame, such that $n^\mu v_\mu = 0$. 

\section{Slow motions}

In this section we will be interested in situations in which the magnitude of peculiar velocity is small compared to the speed of light, i.e. where $v^{\mu} v_{\mu} \ll 1$. 

\subsection{Special case: $v^\mu = 0$ is FLRW} 

Let us start this section by proving that $v^\mu = 0$ corresponds to an FLRW model, so that we can consider the norm of the velocity field, $v^\mu v_\mu$, to be a measure of the departure from an FLRW model. To show this, we note that the limit $v^\mu \rightarrow 0$ implies that the space-time is conformally flat. This can be seen from the vanishing of the 4-acceleration in the frame of the dust source, which implies that $a^\mu=0$ in the quasi-Newtonian frame. This immediately implies from Equation \eqref{sheardot1} that the electric part of the Weyl tensor is zero, which together with the vanishing of the magnetic part from Equation \eqref{magnetic}, means that the Weyl tensor in totality is zero, and thus that the space-time is conformally flat. 
Furthermore, in this case the momentum constraint and the second Bianchi identities imply
\bea
\label{constraints1}
\hspace{-0.2cm}  D_\mu \theta = 0  \qquad {\rm and} \qquad D_\mu \rho = 0 \, ,
\eea 
which means that the space-time must be expanding at the same rate at all points in space, and that the dust must be of uniform density. It follows from Equation \eqref{spatialRicci1} that the trace-free part of the spatial Ricci tensor vanishes. We furthermore have that $D_\mu {}^{(3)}\! R = 0$, since all of the terms going into Equation \eqref{Rtrace2} (note that $T_{\mu \nu}n^\mu n^\nu = \rho$, because the quasi-Newtonian frame and the dust frame coincide) are spatially uniform; cf. Equation \eqref{constraints1}. The 3-dimensional spaces orthogonal to $n^\mu$ are therefore maximally symmetric. It follows from this, and the fact that $n^\mu$ is without rotation or acceleration, that the geometry of the 4-dimensional space-time must be isometric to the spatially homogeneous and isotropic FLRW geometry \cite{Ellis:2011pi}.

\subsection{Linearized equations in $v^\mu$} 

If $v^\mu v_\mu \ll 1$ then we may treat the velocity between matter and quasi-Newtonian frames as an expansion parameter, such that every factor of $v^\mu$ contributes an order-of-smallness\footnote{Note that as $v^\mu$ is the only field that we expand in, the setting is broader than both conventional cosmological perturbation theory and post-Newtonian expansions around an FLRW background.}.
The leading-order contributions to the kinematics of the dust frame can be then found from Equation (\ref{udecomp}) to be given by 
\bea
\label{dynamicsfluid}
\tilde{\theta} =  \theta + D_\mu v^\mu \, , \quad  \tilde{\sig}_{\mu \nu} = D_{\la \mu} v_{\nu \ra} \, , \quad  \tilde{\omega}_{\mu \nu} = D_{[ \nu} v_{\mu ]} \, ,
\eea  
while the leading-order contribution to the 4--acceleration of the quasi-Newtonian frame yields
\bea
\label{acc4}
a^\mu = - \frac{1}{3} \theta v^\mu  -  h^{\mu}_{\, \sigma}  n^\nu \nabla_\nu v^\sigma   - h^{\mu}_{\, \sigma} v^\nu \nabla_\nu v^\sigma  \,  , 
\eea  
where we have kept terms to lowest order in $v^\mu$ only, while allowing its derivatives to be non-perturbative.  We also find that the energy-momentum tensor (\ref{T}) can be decomposed with respect to $n^{\mu}$ as 
\bea
\label{Tn}
\hspace{-0.4cm} T_{ \mu  \nu } = n_{ \mu } n_{\nu } \rho + ( n_{ \mu } h_{ \sigma  \nu }   + n_{ \nu } h_{ \sigma  \mu } ) \rho v^{\sigma}  \, .
\eea
The Raychaudhuri equation in the quasi-Newtonian foliation therefore reads
\bea
\label{raychaudhurin}    
\hspace*{-0.8cm}  n^\mu \nabla_\mu \theta  &=&  -\frac{1}{3}\,\theta^{2} - 4\pi G \rho + \Lambda  + D_\mu a^\mu   \, ,
\eea  
while the evolution equations for the shear of $n^{\mu}$ yield 
\bea
\label{sheardot}
\hspace*{-0.8cm}  
 0 =  -   E_{\mu \nu }  + D_{\la \mu} a_{\nu \ra}  \, ,
\eea 
and the momentum constraint equations give
\bea
\label{constraints}
\hspace{-0.2cm}  D_\mu \theta = 12 \pi G \rho v_\mu \, , \quad \;   D_\nu E_{\mu}^{ \nu } = \frac{8}{3}  \pi G ( D_\mu \rho -  \theta \rho v_\mu ) \, . 
\eea  
It can be seen from Equations \eqref{magnetic} and \eqref{sheardot} that the space-time is conformally flat when $D_{\la \mu} a_{\nu \ra}=0$. The Ricci scalar associated with the 3-dimensional spaces of the quasi-Newtonian foliation is given by Equations \eqref{Rtrace2} and \eqref{Tn}, and the trace-free part is given by \eqref{spatialRicci1} and \eqref{sheardot}, yielding
\bea
\label{spatialRicci}
\hspace*{-0.8cm}  
{}^{(3)}\! R_{\la \mu \nu \ra} = D_{\la \mu} a_{\nu \ra}    \, ,  
\eea 
where we have used the results that shear and anisotropic stress vanish in the quasi-Newtonian frame. Defining ${}^{(3)} G_{\mu \nu}  \coloneqq  {}^{(3)} \! R_{\la \mu \nu \ra} - \frac{1}{6}  {}^{(3)}\! R h_{\mu \nu}$ then gives the conservation equation
\bea
\label{Gconserved}
\hspace*{-0.8cm}  
D_\nu {}^{(3)}  G_{\mu}^{\nu} = 0  \, , 
\eea 
which together with Equation \eqref{spatialRicci} allows us to obtain 
\bea
\label{DHamiltonian} 
 N^2 D_\mu \left( \frac{1}{N^2} \!  {}^{(3)}\! R \right) = 4 D_\mu (D_\nu a^\nu) + 6 a^\nu {}^{(3)}\! R_{\la \mu \nu \ra} \, . 
\eea 
A special case of these equations is considered in Appendix \ref{appendix:A}.

\section{Quasi-Newtonian observables}

Let us consider the class of cosmological observables mediated by congruences of null geodesics with 4-momentum $k^\mu$, and let us suppose (for the moment) that observers and emitters of the light and gravitational waves that follow these congruences are static in the quasi-Newtonian frame (i.e. that they have 4-velocity $n^{\mu}$). This is not a realistic scenario, of course, as we would rather expect observers and emitters to be in the rest frame of the matter. It is, however, computationally convenient as an intermediate step, and can be used to reconstruct observables for observers and emitters in the matter frame with minimal additional effort. 

The redshift, $z$, of a ray of light in the quasi-Newtonian frame is now given by  
\bea
\label{def:zQN}
1+ z  \coloneqq  \frac{E}{E_\obs}   \, , 
\eea 
where $E \coloneqq - n^\mu k_\mu $ is the energy of a photon in this ray{, and the subscript $\mathcal{O}$ indicates that a quantity is evaluated at the observer}. Decomposing the tangent vector to this ray as
\bea
\label{kdecomp}
k^\mu = E (n^\mu - e^\mu) \, , 
\eea 
where $e^\mu$ is a spatial direction vector that is orthogonal to $n^\mu$, such that $n^\mu e_\mu = 0$, we may then write 
\bea
\label{def:zprimeQN}
\hspace*{-0.2cm} \frac{ {\rm d} z}{{\rm d} \lambda} =   - E_\obs (1+z)^2   \left( \frac{1}{3} \theta - e^\mu a_\mu \right)  \, ,   
\eea 
where we have made use of the kinematical identities in and below Equation \eqref{def:nablan}, and where $\lambda$ is an affine parameter along the rays of light. We can directly integrate this equation to get
\bea
\label{def:zintQN}
&& 1   +  z 
= \frac{N_\obs}{N}  \times A_{\rm ISW} \times B_{\rm H} \, , 
\eea 
where
\bea
\label{zAB}
A_{\rm ISW} \coloneqq \text{exp}\left({ - \int_\lambda^{\lambda_\obs} \!\! {\rm d}\lambda' \, E \,    n^\mu \nabla_\mu \ln(N)  }\right)
\quad \text{and} \quad B_{\rm H} \coloneqq \text{exp}\left({ \int_\lambda^{\lambda_\obs} \!\! {\rm d}\lambda' \, E   \, \frac{1}{3}  \theta  }\right) \, . \nonumber
\eea
This result was also found in Reference [\onlinecite{Heinesen:2023lig}], and is particularly useful as the factors within it have clear physical interpretations. The first factor on the right-hand side of Equation \eqref{def:zintQN} is the ratio of lapse functions between the emitter and the observer, and can be interpreted as gravitational redshift. The second factor is given by the integral of the local time-evolution of the lapse function, and can be interpreted as an integrated Sachs--Wolfe effect. The third factor is the cosmological redshift, which comes from the expansion of space along the line of sight (and which reduces to the Hubble expansion in FLRW cosmology); it can be seen to measure the ratio of scales of the quasi-Newtonian spaces by considering the identity 
\be
\label{def:thetadef}
\frac{1}{3} \theta =  n^\mu \nabla_\mu \ln(\mathfrak{a}) = \frac{1}{E} k^\mu \nabla_\mu \ln(\mathfrak{a})  + e^\mu \nabla_\mu \ln(\mathfrak{a})  \, , 
\ee 
which follows from Equations \eqref{def:muint} and \eqref{kdecomp}, and can be used to write  
\bea
\label{def:zintQNchanget2} \nonumber
\hspace{-0.5cm}  && \text{exp}\left({ \int_\lambda^{\lambda_\obs} \!\! {\rm d}\lambda' E \,   \! \frac{1}{3}  \theta  }\right)   = \frac{1}{\mathfrak{a} (\lambda)  }  \text{exp}\left({  \int_\lambda^{\lambda_\obs} \!\! {\rm d}\lambda' E \, e^\mu \nabla_\mu \ln\left( \mathfrak{a} \right)    }\right) .
\eea 
Thus, if the spatial scale $\mathfrak{a}$ is not changing systematically along the spatial direction $e^\mu$, then this redshift factor is (to a good approximation) simply measuring the ratio of spatial scales of quasi-Newtonian spaces between emission and observation. 

Finally, we define the angular diameter distance as 
\bea
\label{def:dA}
d_A \coloneqq \sqrt{ \frac{\delta A }{ \delta \Omega }} \, , 
\eea 
where $\delta A$ is the area of the cross section of the null geodesic congruence orthogonal to $k^\mu$ and $n^\nu$, and $\delta \Omega $ is the angular area of the source in the quasi-Newtonian observer's frame. The evolution equation of $d_A$ is given by the Sachs optical equation \cite{Perlick:2010zh}: 
\be
\label{focuseq}
\frac{{\rm d}^2 d_A }{ {\rm d} \lambda^2} 
= - {\mathcal{F}} d_{A}  \, ,
\ee 
where 
\bea
\mathcal{F} \coloneqq  \frac{1}{2}k^{\mu}k^\nu R_{\mu \nu}  + \frac{1}{2} \hsig^{\mu \nu} \hsig_{\mu \nu} \qquad \text{and} \qquad
\hsig_{\mu \nu} \coloneqq p^{\alpha}_\mu  p^{\beta}_\nu \nabla_{( \alpha } k_{\beta )} - \frac{1}{2}  \nabla_{\alpha } k^{\alpha}  p_{\mu \nu} \, ,\nonumber
\eea
such that $\hsig_{\mu\nu}$ is the shear tensor of the null geodesic congruence and $p^{\alpha}_\mu\coloneqq h^{\alpha}_\mu  - e^\alpha e_\mu$ is the projection operator onto the screen space. We have from Equation \eqref{Tn} that 
\be
\label{focusterm}
\mathcal{F} =  \frac{1}{2} E^2  8 \pi G  \rho + E^2 e^\mu v_\mu  8 \pi G \rho  + \frac{1}{2} \hsig^{\mu \nu} \hsig_{\mu \nu}   \, , 
\ee 
where the first term is the Ricci focusing due to the rest-mass energy density of matter in the quasi-Newtonian frame, the second term is the Ricci focusing due to heat flow of matter, and the third term is the Weyl focusing.

Before finishing this section, let us consider a few particularly interesting aspects of these expressions:

\subsection{Spatial curvature in observations}

Using Equation \eqref{def:zprimeQN}, we can define a new parameter along our null rays:
\be
\label{nu} \nonumber
\nu \coloneqq - E_\obs \int_{\lambda_\obs}^{\lambda} \!\!\! {\rm d}  \lambda' (1+z)^2    =    \int_{\lambda_\obs}^{\lambda} \!\!\! {\rm d} \lambda' \frac{{\rm d} z }{ {\rm d} \lambda'}    \frac{1}{ \left( \frac{1}{3} \theta - e^\mu a_\mu \right) }    \,  ,
\ee 
where we note that the inverse function of $\nu(\lambda)$ is well defined, whereas this is generally not the case for $z(\lambda)$. This allows us to recast Equation \eqref{focuseq} in the form
\be
\label{focuseqnu} 
 \hspace{-0.1cm}
\frac{{\rm d}^2 (1+z) d_A }{ {\rm d} \nu^2} 
= - \kappa (1+z) d_{A} \, , 
\ee  
where using Equations \eqref{Tn}, \eqref{raychaudhurin} and \eqref{Rtrace2} we find
\bea
\label{kappa} 
\kappa \coloneqq  \frac{ \mathcal{F} + E^2  \frac{1}{E} \! \frac{{\rm d}  \left( \frac{1}{3} \theta - e^\mu a_\mu \right)  }{ {\rm d} \lambda }  }{E_\obs^2 (1+z)^4   }
=\frac{  \frac{1}{6}  {}^{(3)}\! R + \frac{D_\mu a^\mu}{3} - \frac{1}{E^2} k^\mu k^\nu \nabla_{\mu} a_\nu + \frac{1}{2 E^2} \hat{\sigma}^{\mu\nu} \hat{\sigma}_{\mu\nu}    }{ (1+z)^2   } \, .
\eea
In the FLRW limit, $\kappa$ is constant and equal to the spatial curvature parameter $k_{\text{FLRW}}$, such that the solution to Equation \eqref{focuseqnu} yields the familiar expressions from Friedmann cosmology. In the general quasi-Newtonian case, the numerator in the expression for $\kappa$ above reduces to the homogeneous Friedmann curvature when $a^\mu = 0 $, as can be seen from Equation \eqref{Rsol} in Appendix \ref{appendix:A} and by using the equation 
\bea
\label{def:shearphotons}
\hspace{-0.5cm}  k^\alpha \nabla_\alpha \hat{\sigma}_{\mu\nu} = - \hat{\theta} \hat{\sigma}_{\mu\nu}  -2 E^2 p^{\alpha}_{\, \la \mu} p^{\beta}_{\,  \nu \ra} D_\alpha a_\beta \, , 
\eea 
which can be deduced from Equations \eqref{magnetic} and \eqref{sheardot}. We note that it follows from quite general considerations of light beams near focus points that $\hat{\sigma}_{\mu\nu}=0$ must be zero at the vertex, and is subsequently zero throughout, if the Weyl curvature of the space-time vanishes \cite{Seitz:1994xf}. In the present case, it can be seen that light cones are shear-free if $D_{\alpha} a_{\beta}$ vanishes. Thus, all departures of $\kappa$ from FLRW expectations are due to non-zero (gradients of) $a^\mu$. Departures from an FLRW space-time could be assessed observationally by probing the non-constant nature of $\kappa$, which is the basic idea behind the FLRW test devised by Clarkson, Basset and Lu \cite{PhysRevLett.101.011301}. 

\subsection{Returning to the matter frame}
\label{sec:obsmatter}

It is now of interest to transform the motion of observers and emitters so that they are in the matter frame, rather than being stationary in the quasi-Newtonian spaces. Starting with the redshift, we can define this in the matter frame by
\bea
\label{def:zdustrans}
\hspace*{-0.45cm} 1 + \tilde{z} \coloneqq \frac{\tilde{E}}{\tilde{E}_\obs}=    (1+z) \frac{\gamma(1 + e^\mu v_\mu)}{\gamma_{\mathcal{O}}(1 + e^\mu v_\mu \rvert_{\obs} \!)}  \, , 
\eea 
where $\tilde{E} \coloneqq - u^\mu k_\mu$ is the energy of a photon as measured by an observer comoving with matter, and in the second equality we have used Equations \eqref{udecomp} and \eqref{kdecomp}, which relate the relevant quantities in the two frames. The additional factors on the right-hand side of this equation can be understood simply as the Doppler shift due to the boost between frames at the location of the observer and the emitter.

We further note that the angular diameter distance in the matter frame will be defined as
\bea
\label{def:dAdust}
\tilde{d}_A \coloneqq \sqrt{ \frac{ \delta \tilde{A} }{ \delta \tilde{\Omega} }} \, , 
\eea 
which we find to be related to the corresponding expression in the quasi-Newtonian frame by 
\bea
\label{def:dAdusttrans}
\tilde{d}_A  =   \frac{ \tilde{E}_\obs }{ E_\obs }   d_A =  \gamma_{\mathcal{O}} (1 +  e^\mu v_\mu \rvert_{\obs} \!)  \, d_A    \, , 
\eea 
where the first equality follows from the aberration of angles through the transformation of the Jacobi map (see Equation~2.26 in Reference [\onlinecite{Korzynski:2017nas}]), and the last equality again follows from Equations \eqref{udecomp} and \eqref{kdecomp}. This completes the transformation rules of our observables to the matter frame.

\section{Example: The Quasi-Newtonian frame in Kasner space}
\label{sec:ObsKasner}
We shall take as an example the degenerate Kasner solution of Einstein's equations, which has $T_{\mu\nu} =0 = \Lambda$, and is described by the following line element:
\begin{equation}
    \label{eq:Kasner}
    \begin{aligned}
    ds^2 &= -d{\tC}^2 + {\tC}^{-\frac{2}{3}}dx^2 + {\tC}^{\frac{4}{3}}dy^2 + {\tC}^{\frac{4}{3}}dz^2 \, .
    \end{aligned}
\end{equation}
This space-time is clearly quite different from FLRW when described by kinematic quantities defined by the time-like curves along which the spatial coordinates are constant: it has non-zero shear and a non-zero Weyl tensor. Nevertheless, we will show that it admits a quasi-Newtonian description in which the observational relationships derived above remain valid.

As this is a vacuum space-time, there is no unique matter frame. We therefore choose to take the frame defined by the time-like 4-velocity $u_{\mu}= -\nabla_{\mu}{\tC}$ to play this role, and will refer to it as the `canonical frame'. The integral curves of this 4-vector field are orthogonal to hypersurfaces of spatial homogeneity, and so are geometrically special. They exhibit vanishing vorticity, but non-vanishing shear. Performing a boost in the $x$-direction we find that Equation (\ref{udecomp}) allows us to write
\begin{equation}
    \label{eq:v}
    v^{\mu} = -\gamma \bigg(\frac{v^{2}}{{\tC}^{\frac{2}{3}}} \, \delta^{\mu}_{\;0}+ v\,\delta^{\mu}_{\;1}\bigg)\,,
\end{equation}
where $\gamma=(1-v_{\mu} v^{\mu})^{-{1}/{2}}=(1-{v^{2}}{{\tC}^{-2/3}})^{-{1}/{2}}$ is the relevant Lorentz factor. This boost does not induce any vorticity, and can be shown to lead to a shear-free congruence, $n^{\mu}$, of time-like curves, if the following equation is obeyed:
\begin{equation}
\begin{aligned}
	\label{eq:shearfreeconditionK}
	2v^2 - 3{\tC}^{2/3}(1-{\tC}\,v_{,x})+ 3{\tC}\,v \, v_{,{\tC}}=0 \, .
\end{aligned}
\end{equation}
It follows that, to leading order in $v$, the boost required to reach the canonical frame at every point in space-time is given by Equation (\ref{eq:v}) with 
\begin{equation}
    \label{eq:QNv_expansion}
    v(x,{\tC}\,) = \frac{x}{{\tC}} + O(x^3) \, ,
\end{equation}
where we have chosen to expand around the point $x=0$, which we take to be the location of the observer.
We note that the slow-motion limit, $v \ll 1$ naturally imposes a local domain of validity for our approximation. As $x$ approaches $\tC$ the quasi-Newtonian spaces become increasingly deformed relative to the canonical Kasner spaces. We expect this to continue until $v$ becomes super-luminal and the foliation breaks down, which (if true) means that to construct a quasi-Newtonian foliation in the degenerate Kasner space one must remain within a suitable spatial domain around the origin (such that $x \lesssim \tC$). Further, to satisfy $v \ll 1$ we clearly require $x \ll \tC$, which corresponds to remaining within an even smaller region of space (one much smaller than the Hubble radius). Finally, we note that expanding our expressions consistently to some order in $v$ will require us to treat spatial and temporal derivatives on a different footing, such that spatial derivatives remove an order-of-smallness. This is clear from Equation \eqref{eq:QNv_expansion}, and is in keeping with the way in which derivatives are treated in post-Newtonian expansion (see, e.g., Reference [\onlinecite{poisson2014gravity}]).

\subsection{Observational relations in Kasner}
\label{subsec:obsKasner}

Let us begin by calculating the redshift between emitter and observer using Equation (\ref{def:zintQN}), for photons that travel in the $x$-direction. For this we need the lapse function $N$, which from Equation (\ref{gradient}), and using the integrability condition $\tQN_{,\tC x}=\tQN_{,x \tC}$, we find must obey
\begin{equation}
    \label{eq:integrable}
    N_{,{\tC}} \bigg(\frac{\gamma\, v}{{\tC}^{{2}/{3}}}\bigg) + N_{,x}\gamma - N\bigg(  \frac{\gamma\, v}{{\tC}^{{2}/{3}}}\bigg)_{,{\tC}} - N \gamma_{,x}=0 \, ,
\end{equation}
which admits the following solutions for lapse and $\tQN$
\begin{equation}
    \label{eq:lapse}
    N = \frac{\gamma}{h\bigg(\frac{{\tC}}{\gamma}\bigg)_{,{\tC}}} \qquad {\rm and } \qquad \tQN = h\bigg(\frac{{\tC}}{\gamma}\bigg) \, ,
\end{equation}
where $h$ is some monotonic function. We can constrain this solution by imposing that ${\tC} = \tQN $ at the location of the observer, which implies 
\begin{equation}
    \label{eq:laspeexpand}
    N = 1 +{\tC}^{1/3} v_{,{\tC}}\,v + \frac{2v^2}{3{\tC}^{2/3}} +O(v^3) \, .
\end{equation}
Now, using Equations (\ref{udecomp}) and (\ref{kdecomp}), we may write
\begin{equation}
    \label{eq:tangentvector}
    \frac{{\rm d}{\tC}}{ {\rm d}\lambda} = -u_{\mu}k^{\mu}  = E(\gamma + e^{\mu}u_{\mu}) = E\gamma\bigg(1 - \frac{v}{{\tC}^{1/3}}\bigg)\, ,
\end{equation}
where in the last equality we have used the Lorentz transformation between canonical and quasi-Newtonian frames, which gives $e^{\mu} = \gamma(v {\tC}^{-1/3}\,\delta^{\mu}_{\;0} + {\tC}^{1/3}\,\delta^{\mu}_{\;1})$. Putting these results into Equation (\ref{def:zintQN}) then gives the redshift
\bea \label{eq:QNredshiftKasner}
1+z = \bigg(1 + \frac{v^2}{3{\tC}^{2/3}} + {\scriptstyle \ldots}\bigg) \times A \times B \, ,
\eea
where
\bea \nonumber
&&A \coloneqq \exp{\bigg(\!\! -\int^{{\tC}_{\mathcal{O}}}_{{\tC}} \!\!\!{\rm d}{\tC}' \bigg(\frac{2v^{2}}{9{\tC}'^{\frac{5}{3}}}+ {\scriptstyle \ldots}\bigg) \!\!\bigg(1 \!+\! \frac{v}{\tC'^{\frac{1}{3}}} \!+\! \frac{v^{2}}{2{\tC}'^{\frac{2}{3}}}\!+\! {\scriptstyle \ldots}\!\bigg)\!\!\bigg)}\,,\\
&&B \coloneqq \exp{\bigg(\! \int^{{\tC}_{\mathcal{O}}}_{{\tC}}\!\!\!{\rm d}{\tC}' \bigg(1 \!+\! \frac{v}{{\tC}'^{\frac{1}{3}}} \!+\! \frac{v^{2}}{2{\tC}'^{\frac{2}{3}}}\!+\! {\scriptstyle \ldots}\!\bigg)\frac{1}{3}\bigg(\frac{2}{{\tC}'} +\frac{v^2}{{\tC}'^{\frac{5}{3}}}+{\scriptstyle \ldots}\bigg)\!\!\bigg)} \, ,
\eea
and where we have used Equation (\ref{eq:QNv_expansion}) and the fact that $\theta\coloneqq\nabla_{\mu}n^{\mu}= {2}\gamma {\tC^{-1}}  =2\tC^{-1} + {v^{2}}{\tC^{-5/3}}+O(v^3)$. We note that the leading-order term in the expansion of $\nabla_{\mu} n^{\mu}$ is identical to the expansion of a comoving observer in an Einstein-de Sitter universe, while the next-to-leading-order term would usually be attributable to an inhomogeneous perturbation of that space-time.

From the equations of motions of the photons following an $x$-directed null geodesic we find, in Appendix \ref{appendix:B}, that
\begin{equation}
    \label{eq:x(t)_null}
    x({\tC}\,) = -\frac{3}{4}\bigg({\tC}^{\,4/3} - {\tC}^{\,4/3}_{\mathcal{O}}\bigg)
\end{equation}
where the constant $\tC_{\mathcal{O}}$ is the time of observation for the observer at $x=0$. This allows us to write Equation (\ref{eq:QNredshiftKasner}) as
\begin{equation}
    \label{eq:QNKasnerRedshiftResult}
    1+z = 1 + \frac{2}{3}\bigg(\frac{{\tC}_{\mathcal{O}} - {\tC}}{{\tC}}\bigg) + \frac{5}{9}\bigg(\frac{{\tC}_{\mathcal{O}} - {\tC}}{{\tC}}\bigg)^2 +O\left(\bigg(\frac{{\tC}_{\mathcal{O}} - {\tC}}{{\tC}}\bigg)^3\right)\,, 
\end{equation}
where we have used $({\tC}_{\mathcal{O}} - {\tC})\,{\tC}^{-1} \ll1$, which follows from $v \ll1$. Boosting the observer and the emitter to the canonical frame, using the transformation rule in Equation (\ref{def:zdustrans}), then gives
\begin{equation}
    \label{eq:matterKasnerRedshiftResult}
    1 +\tilde{z} =  1 - \frac{1}{3}\bigg(\frac{{\tC}_{\mathcal{O}} - {\tC}}{{\tC}}\bigg) +  \frac{2}{9}\bigg(\frac{{\tC}_{\mathcal{O}} - {\tC}}{{\tC}}\bigg)^2 +O\left(\bigg(\frac{{\tC}_{\mathcal{O}} - {\tC}}{{\tC}}\bigg)^3\right)\, ,
\end{equation}
where $ e^\mu v_\mu =0$ at the observer located at $x=0$. This result can be seen to precisely reproduce the known result for redshift in this space-time, as given by Equation \eqref{zkasner} in Appendix \ref{appendix:B}, up to second order in $({\tC}_{\mathcal{O}} - {\tC})\,{\tC}^{-1}$.

To calculate the angular diameter distance we note that null rays propagating in the $x$-direction of the degenerate Kasner metric (\ref{eq:Kasner}) must be shear-free. Together with the fact that this is a vacuum space-time, it can be seen that the right-hand side of Equation (\ref{focuseq}) vanishes, so
\begin{equation}
    \label{eq:daQN}
    \frac{{\rm d} {d}_A }{ {\rm d} \lambda } = -E_{\mathcal{O}} \,,
\end{equation}
where the integration constant $E_{\mathcal{O}}$ is fixed by the initial conditions for ${{\rm d} {d}_A }/{ {\rm d} \lambda }$ at $\lambda=\lambda_{\mathcal{O}}$. {As we have chosen the quasi-Newtonian and canonical frames to coincide at the observer, Equation (\ref{def:dAdusttrans}) shows that $\tilde{d}_A=d_A$. From this, and using Equations (\ref{def:zQN}) and (\ref{eq:tangentvector}) we find}
\bea \nonumber
    \tilde{d}_A &=& \int^{{\tC}_{\mathcal{O}}}_{{\tC}} \frac{{\rm d}{\tC}\,'}{\gamma(1+z)\bigg(1 -\frac{v}{{\tC}^{1/3}}\bigg)}
   \simeq  \frac{1}{\mathcal{H}_{\mathcal{O}}} \left[-\frac{1}{3} \bigg( \frac{{\tC}_{\mathcal{O}} - {\tC}}{{\tC}} \bigg) + \frac{5}{18}\bigg( \frac{{\tC}_{\mathcal{O}} - {\tC}}{{\tC}} \bigg)^2 \right] {\simeq \frac{1}{\mathcal{H}_0}\,\left(\tilde{z} + \frac{1}{2}\tilde{z}^2\right) } \, ,   
   \label{eq:d_Aintegral}
\eea
where $\mathcal{H}_{\mathcal{O}}= - \frac{1}{3} \tC_{\mathcal{O}}^{-1}$ is the expansion of space in the $x$-direction in the canonical frame, and where we have used Equation \eqref{eq:matterKasnerRedshiftResult} to change variable to $\tilde{z}$. This perfectly reproduces the exact expression for $\tilde{d}_A(\tilde{z})$ in Equation \eqref{dAkasner} of Appendix \ref{appendix:B}.

Having demonstrated that Equations \eqref{def:zintQN} and \eqref{focuseq} correct reproduce the expected results in the degenerate Kasner geometry \eqref{eq:Kasner}, let us turn to the measures of spatial curvature in the quasi-Newtonian foliation, given by Equations \eqref{Rtrace2} and \eqref{kappa}. For a vacuum space-time with zero cosmological constant we find that the three-dimensional Ricci curvature scalar in this case is given by
\begin{equation}
    \label{eq:QNspatialRicciScalar}
    ^{(3)}R = -\frac{8}{3{\tC}^2}\gamma^2\,,
\end{equation}
which is everywhere negative. Similarly, we may compute $\kappa$, as defined in Equation $(\ref{kappa})$. Noting that $\mathcal{F}=0$ for a bundle of shear-free null geodesics propagating in the $x$-direction, using Equations (\ref{def:zQN}) and (\ref{eq:tangentvector}) we find this spatial-curvature-like term from the focusing equation is given by
\bea
    \label{eq:QNKkappa}
    \kappa &=& (1+z)^{-2} (\gamma + e^{\mu}u_{\mu}) \frac{\rm{d}}{\rm{d}{t}}\bigg( \frac{1}{3}\theta - e^{\mu}a_{\mu}\bigg) 
    \simeq -(1+z)^{-2}\bigg( \frac{2}{3{\tC}^2} + \frac{8 v}{9{\tC}^{7/3}}  \bigg) \, .
\eea 
In a perfect FLRW geometry this function would be constant and equivalent to the usual FLRW spatial curvature parameter $k$.  In the present case it can be seen to be negative close to the observer, in the region where the perturbative expansion is valid.   
It is interesting that both the kinematics and the optical properties of this space-time indicate negative spatial curvature in the quasi-Newtonian description, as the Kasner space-time itself is in fact Bianchi type I, and is therefore usually thought of as being spatially flat (in the canonical description). 

\section{Implications for the real Universe and for Cosmological modeling}

It seems plausible that, at least over some range of temporal and spatial scales, there should exist in the real Universe an approximately quasi-Newtonian frame of the type investigated in this paper. The presence of large amplitude gravitational waves, frame dragging effects, or highly relativistic matter (e.g. neutrinos) would spoil such a conclusion, but as long as these phenomena are sub-dominant gravitationally then we would expect to be able to include them as small perturbations to what would otherwise be a perfect shear-free description of space. Even the Kasner solution that we considered as an example, which exhibits strong anisotropy and has shear of order unity in its canonical foliation, can be mapped to a quasi-Newtonian description within domain sizes that are smaller than the Hubble horizon. We therefore find it plausible that the formalism investigated in this paper represents a realistic way to model at least parts of the Universe, independent of the precise cosmological model used to describe it globally.

In the quasi-Newtonian description, departures from the FLRW distance-redshift relation are controlled by the acceleration field $a^\mu$. If this field should happen to cancel along rays of light over long distances, because e.g. all of the relevant fields are distributed in a regular way around zero, then we find that observations in the quasi-Newtonian frame should be well described by the FLRW laws governing optical measures of distance and redshift. In addition, we find that if the velocity field of the matter in the Universe, $v^\mu$,  is of small amplitude, as well as being randomly distributed around zero, then the transformations in Equations \eqref{def:zdustrans} and \eqref{def:dAdusttrans} should have little average effects on observables, apart from dipole contributions coming from the motion of the observer and emitter. 

On the other hand, if either (i) there exists a setting where $a^\mu$ or its gradients have persistent sign or orientation, such that their contributions do not cancel along lines of sight, and/or (ii) there is a persistent flow $v^\mu$ of the matter in the Universe relative to the quasi-Newtonian frame, then we do not expect to recover the standard FLRW relations. The degenerate Kasner solution that we considered is an example of such a space-time, where $a^\mu$ and $v^\mu$ have persistent orientation in the $x$-direction. Conditions (i) and (ii) are related through Equation \eqref{acc4}, and we find it likely that cosmological models that exhibit one of the conditions will very likely also exhibit the other. Such models would then have highly non-trivial kinematics and observables on cosmological scales, and would therefore be of great interest. We note that this could apply not only to cosmological models that violate the cosmological principle, which would certainly have persistent velocities, but potentially also to those that are statistically homogeneous in the matter frame, but which could have a rest mass density with persistent non-zero gradients in the quasi-Newtonian frame. 

This approach provides us with an interesting way to understand a number of different types of cosmological models, and the observations that may be made within them. For example, it is clear that cosmological models that are built from perturbations around an FLRW background must stay close to an FLRW universe in terms of both kinematics and observables, at least if the perturbations contain only sub-horizon modes. Furthermore, in most post-Newtonian approaches for describing structures with large density contrasts the velocity field and acceleration field remain small (by construction) and without persistent orientations when the structures are modest in size, thus resulting in Friedmannian light propagation \cite{clifton2009hubble}. This is also true of Swiss cheese models with embedded LTB and Szekeres structures \cite{clifton2009hubble, Lavinto:2013exa,fleury2014swiss, Koksbang:2017arw}, which exhibit non-perturbative local rates of differential expansion and shear in the matter frame, but tend to behave very simply globally. Within the context of our present study this can be understood to be due to the acceleration and velocity fields vanishing in between structures in a quasi-Newtonian foliation, such that there are no persistent gradients of $a^\mu$ or $v^\mu$ on cosmological scales. The average of kinematics and observations in such models must then conform well to those of an FLRW model, which is consistent with the idea that small enough regions of LTB (and presumably non-singular Szekeres) should be well modeled by Newtonian theory \cite{VanAcoleyen:2008cy}.

It is less obvious that cosmological models with non-perturbative structures should in general have optical properties similar to FLRW models. Indeed, we expect that this should not generically be what happens in general relativistic cosmological space-times, even if the space-times obeys a statistical version of the cosmological principle \cite{BuchertRasanen}. However, there is a lack of concrete realizations of this idea in the literature, because the scenarios used to investigate the phenomenon, such as those discussed above, typically (implicitly) impose that persistent $a^{\mu}$ and $v^{\mu}$ fields cannot emerge. The approach taken in this paper offers a possible resolution to this bottleneck in cosmological modeling, by providing a clear statement of the conditions necessary for non-trivial behaviors to emerge over large scales in the Universe.
 
\section{Conclusions} 
\label{sec:conclusion} 

We have considered observational relationships in quasi-Newtonian foliations of arbitrary cosmological models, and produced conditions for convergence to the results of FLRW models. We find that the cosmological redshift can be expressed in terms of a Hubble expansion contribution, an integrated Sachs-Wolfe effect, and a gravitational redshift contribution, as shown in Equation \eqref{def:zintQN}. This is a decomposition that is familiar from standard cosmological perturbation theory around FLRW backgrounds, and which persists in the quasi-Newtonian foliation of more general models. The focusing of the angular diameter distance is also found to be expressible in terms of  the familiar mass density, heat flow, and Weyl focusing contributions \eqref{focusterm}. We have illustrated these ideas with a simple non-FLRW model; the degenerate Kasner solution. The impact of anisotropic expansion of space on light-propagation in this geometry can be understood in terms of the familiar quantities from cosmological perturbation theory around the isotropically expanding Friedmann solutions, once a transformation to the quasi-Newtonian frame is made. While we expect that in a fully relativistic setting it will, in general, not be possible to identify a hypersurface forming congruence of vanishing shear, we do expect frames of minimal shear to exist. These have been studied in the numerical context as a generalization of the linearized Poisson gauge \cite{Giblin:2018ndw}, and are assumed to exist from the outset in the general relativistic N-body code {\it gevolution} \cite{Adamek:2016zes} (up to gravitational waves and frame dragging fields). We find it interesting that, within the particular example of the Kasner space-time we consider, the shear-free condition seems to naturally lead to a hierarchy of size in temporal and spatial derivatives, such that a post-Newtonian expansion around an FLRW background  naturally emerges. It would be interesting to see if this is a property that emerges for broader classes of space-times, as it would further add to the list of Newtonian-like properties of shear-free foliations. 

We consider the results of our investigation to also be interesting in light of the present cosmological tensions that feature internal inconsistencies in cosmological parameters describing the average kinematics of the standard model \cite{di2021realm}, but also anomalies in bulk flows \cite{watkins2023analysing} and in dipolar features in number counts \cite{2021ApJ...908L..51S} and distance moduli \cite{migkas2025galaxy}. It appears to be a generic feature of the quasi-Newtonian descriptions considered here that deviations in the average kinematics should be directly related to persistent peculiar velocity fields across large cosmic distances. We find it an intriguing idea that seemingly unrelated tensions in cosmology could have a common origin, but make no attempt here to propose a concrete mechanism or model that could directly explain these anomalies. We do, however, hope that our results provide future avenues of research for these problems.

\begin{acknowledgments}
This work was supported by the Frederick Perren Fund and the STFC Research Council.
\end{acknowledgments}








\newpage

\appendix

\section{Exact integration in simplified case}
\label{appendix:A}

Let us consider the case where $\abs{ a^\nu {}^{(3)}\! R_{\la \mu \nu \ra} } = \abs{ a^\nu D_{\la \mu} a_{\nu \ra} }   \ll \abs{ D_\mu (D_\nu a^\nu)}$, which is true in both cosmological perturbation theory and post-Newtonian theory. We may then neglect the second term in Equation \eqref{DHamiltonian},  and integrate its leading-order part to yield 
\bea
\label{Rsol} 
 {}^{(3)}\! R =  N^2 \mathcal{R}^{\text{hom}} + 4 D_\nu a^\nu \, , 
\eea 
where $\mathcal{R}^{\text{hom}}$ is a homogeneous function satisfying $D_\mu \mathcal{R}^{\text{hom}} = 0$. Using the Equation \eqref{vorticitydot} in Equations \eqref{spatialRicci} and \eqref{Rsol}, and comparing to the formulae in Equations \eqref{Rtrace} and \eqref{Rtracefree}, we find that a solution for $\mu$ is given by
\bea
\label{Solutionmu} 
\hspace*{-0.5cm}  \frac{\mu^{\frac{1}{3}} }{ \mu^{\frac{1}{3}}_{\text{i}}  } = \frac{\hat{\mathfrak{a}}(\tQN)}{ \hat{\mathfrak{a}}(\tQN_{\text{i}})} \frac{N_{\text{i}} }{N}  \, 
\eea 
when the following is simultaneously satisfied:
\bea
\label{solmucondition} 
&& \mathcal{R}^{\text{hom}} =  \frac{ \mathcal{R}^{\text{hom}}_{\text{i}} \hat{\mathfrak{a}}^2(\tQN_{\text{i}})  }{  \hat{\mathfrak{a}}^2(\tQN) } \, , 
\eea 
where $\hat{\mathfrak{a}}(\tQN)$ is an arbitrary function in $\tQN$. The subscript \sayy{$\text{i}$} here denotes evaluation at an initial time $\tQN_{\text{i}}$. The solution for the volume element in Equation \eqref{Solutionmu} takes a similar form as in the Poisson gauge of cosmological perturbation theory as well as in Newtonian expansions around an FLRW background. For this solution, we furthermore have that 
\bea
\label{solthetacondition} 
&& \theta = \frac{n^\nu \nabla_\nu \mu }{\mu} = 3 \left( \frac{\partial_{\tQN} \hat{\mathfrak{a}} }{\hat{\mathfrak{a}}} -   \frac{\partial_{\tQN} N }{N} \right) . 
\eea 
In this case, any spatial variation in $\theta$ comes from the time evolution of the lapse function. In the case of static potentials, we then recover the results of Buchert and Ehlers, that back-reaction vanishes in Newtonian cosmologies up to boundary terms \cite{buchert1995averaging}. We finally remark that the solution in Equation \eqref{Rsol}, when integrated over spatial domains orthogonal to $n^\mu$, reduces to a component with the FLRW scaling law in Equation \eqref{solmucondition} (from the first term) and a boundary term (from the second term), when applying Gauss's theorem. This exemplifies the role that boundary conditions play for average dynamics in the weak-field limit of cosmology. 

\section{Exact observational relations in Kasner}
\label{appendix:B}

Solving the Euler-Lagrange equations for light rays moving along the x-direction, in a space-time described by Equation~\eqref{eq:Kasner}, we find that
\bea
\label{nullpath}
k^{\mu}\coloneqq \frac{\rm{d}x^{\mu}}{\rm{d}\lambda}=\left( \frac{\tilde{E}_{\mathcal{O}}}{\tC_{\mathcal{O}}^{\frac{1}{3}}}\; \tC^{\frac{1}{3}}\,,\; \pm \frac{\tilde{E}_{\mathcal{O}}}{\tC_{\mathcal{O}}^{\frac{1}{3}}}\; \tC^{\frac{2}{3}}\,  , \;0\,\;0\right)
\eea 
where the constant of integration is fixed by requiring that an observer in the canonical frame measures the energy of the incoming photons to be $\tilde{E}_{\mathcal{O}}$. From this, and Equation \eqref{def:zdustrans}, we find the redshift to be given by the exact relation
\bea \label{zkasner}
1 + \tilde{z} =  \left(\frac{\tC}{{\tC}_{\mathcal{O}}}\right)^{\frac{1}{3}}\,.
\eea
Recalling that Equation \eqref{focuseq} is valid in the canonical frame, and that its right-hand side is zero, we find that Equation \eqref{nullpath} gives
\bea \label{dAz}
\tilde{d}_{A} = \tilde{E}_{\mathcal{O}}(\lambda_{\mathcal{O}}-\lambda)= \tC_{\mathcal{O}}^{\frac{1}{3}} \int^{\tC_{\mathcal{O}}}_{\tC} d\tC'\tC'^{-\frac{1}{3}} = \frac{3}{2}\tC_{\mathcal{O}} \left(1 - \left(\frac{\tC}{\tC_{\mathcal{O}}}\right)^{\frac{2}{3}} \right)\;.
\eea
Writing the angular diameter distance as a function of redshift, via Equation \eqref{zkasner}, we conclude that
\bea \label{dAkasner}
\tilde{d}_A = \frac{1}{\mathcal{H}_0}\,\left(\tilde{z} + \frac{1}{2}\tilde{z}^2\right)
\eea
where $\mathcal{H}_0=-1/(3\,{\tC}_{\mathcal{O}})$. The relations obtained in Equations \eqref{zkasner} and \eqref{dAkasner} are both exact, and have been derived using standard general relativistic methods.

\section*{References}

\bibliographystyle{unsrt}

\end{document}